# Temperature and excitation energy dependence of Raman scattering in nodal line semimetal ZrAs$_2$

*R. Bacewicz[1], C. Jastrzębski[1], K. Zberecki[1], A. S. Wadge[2], D. Jastrzębski[2] and A. Wiśniewski[2,3]*

[1]*Faculty of Physics, Warsaw University of Technology, Koszykowa 75, PL- 00-662, Warsaw, Poland*
[2]*International Research Centre MagTop, Institute of Physics, Polish Academy of Sciences, Aleja Lotnikow 32/46, PL-02-668 Warsaw, Poland*
[3]*Institute of Physics, Polish Academy of Sciences, Aleja Lotnikow 32/46, PL-02-668 Warsaw, Poland*

ABSTRACT

We present a Raman study of ZrAs$_2$ single crystals, a nodal line semimetal with symmetry-enforced Dirac-like band crossings. We identified the symmetry of phonon modes by polarized light measurements and comparison with calculated phonon frequencies. Significant dependence of peak intensities on the excitation wavelength was observed, indicating quantum interference effects. Phonon peaks in the spectra are superimposed on the electronic background, with quasi-elastic scattering observed for the 785 nm excitation. We identified the Fano shape of the 171 cm$^{-1}$ A$_g$ mode due to interference of the phonon state with the electronic continuum. The temperature dependence of phonon peaks linewidth indicates that the electron-phonon coupling plays an essential role in phonon decay.

INTRODUCTION

Topological semimetals have attracted considerable scientific interest in recent years owing to their unique electronic structure and the specific role of symmetry. In most Dirac and Weyl semimetals, the energy bands intersect at a point in k-space. Nodal line semimetals exhibit band crossings that extend on a one-dimensional line or loop in k-space [1]. The nonsymmorphic symmetries can play an important role in protecting the crossings [2,3]. Transition metal dipnictides ZrP$_2$ [4] and ZrAs$_2$ [5] have been found to represent nodal line semimetals. Bannies et al. [4] using angle-resolved photoemission spectroscopy (ARPES) and magnetotransport studies found that ZrP$_2$ exhibits an extremely large and unsaturated magnetoresistance (MR) of up to 40 000 % at 2 K, which originates from an almost perfect electron-hole compensation. Their band structure calculations and ARPES studies showed that ZrP$_2$ hosts a topological nodal loop in proximity to the Fermi level. Very recently, magnetotransport studies of ZrAs$_2$ have been reported by Nandi et al. [6]. They observed large MR with quadratic field dependence, unsaturated up to magnetic field of 14 T. Their electronic structure analysis demonstrates the coexistence of electron and hole pockets at the Fermi surface. The carrier concentration was estimated from the field-dependent Hall resistivity, and it was found that the charge carriers are nearly compensated, which results in a large MR. Wadge et al. [5], reported results for the $ZrAs_2$ single crystals, obtained using ARPES technique and DFT calculations. In ARPES scans, a distinctive nodal loop structure was observed at lower photon energies of 30 and 50 eV. Furthermore, DFT calculations unveiled symmetry-enforced band crossings anchored at specific points in the Brillouin zone.

Raman scattering studies of topological semimetals offer insight into lattice dynamics, electronic structure, and electron-phonon interaction [7-10]. This paper presents the Raman study of zirconium di-arsenide $ZrAs_2$ single crystals. Since, to our knowledge, no Raman studies on $ZrAs_2$ have been published yet, we performed angle-resolved polarization measurements to assign symmetry to the observed Raman modes. With the support of the phonon frequencies ab initio calculations, it was possible to identify all Raman modes in $ZrAs_2$. The Raman spectra depend substantially on the excitation energy, with some modes visible only for specific excitation. We ascribe this to resonance-like/interference effects related to the complicated band structure of $ZrAs_2$. We also analyzed the electronic Raman scattering, which produces a pronounced continuum background in the Raman spectra. An essential aspect of the Raman spectroscopy of semimetals is the role of electron-phonon coupling. Our study shows the electron-phonon coupling effects in $ZrAs_2$, manifesting in Fano resonance, and the temperature dependence of Raman peaks linewidth.

EXPERIMENTAL DETAILS

$ZrAs_2$ crystallizes in the $PbCl_2$ – type structure with the centrosymmetric, nonsymmorphic space group Pnma ($D_{2h}^{16}$). Orthorhombic cell parameters are [11]: a = 6.8006 Å, b = 3.6883 Å, c = 9.0328(4) Å. The unit cell of $ZrAs_2$ contains 4 formula units (Fig. 1), all atoms occupy 4c Wycoff positions. Needle-like crystals of $ZrAs_2$ have been grown by the iodine transport method, with the crystallographic b axis along the needle length. The chains of covalently bonded As atoms along the b-axis provide preferred direction of crystal growth [11].

Raman measurements were performed on two spectrometers: Horiba Jobin Ivon Aramis spectrometer with 2400 l/mm diffraction grating (appr. resolution 1 $cm^{-1}$), used for measurements with 473 nm and 633 nm excitation lasers; Renishaw Quanta spectrometer with diffraction grating 1800 l/mm and approximate resolution of 2 $cm^{-1}$ (785 nm laser). The beam power was kept low in all cases to prevent sample heating. The spectra were measured in the temperature range 80 – 443 K.

Raman spectra were measured in backscattering geometry with parallel ($\hat{e}_i \parallel \hat{e}_s$) and crossed ($\hat{e}_i \perp \hat{e}_s$) configurations, where $\hat{e}_i$, $\hat{e}_s$ are polarization vectors of incident and scattered light, respectively. For most of the measurements the natural (101) face of the crystal was used, with light impinging along the Z' axis, perpendicular to the (101) face (Fig. 2). We use Porto's notation throughout the paper, e.g. Z'(YY)$\bar{Z}$' configuration, which we write in short as (YY).

COMPUTATIONAL DETAILS

The calculations of the band structure were performed within density functional theory (DFT) as implemented in the VASP Package [12-15] with Projector augmented wave pseudopotentials (PAW) [16, 17]. In all cases, the PBE (GGA) functional [18] has been used. For the sampling of the Brillouin zone, a dense 8 × 8 × 8 grid was used, while the plane wave energy cutoff was set to 500 eV. All the structures were optimized until the force exerted on each atom was smaller than $10^{-5}$ eV/Å. Phonon dispersion calculations were made using, the frozen phonon method as implemented in the phonopy code [19].

RESULTS AND DISCUSSION

**Room temperature studies**

According to the nuclear site group analysis [20] of ZrAs$_2$ structure, 18 Raman active optical vibrational modes at zero wavevector are distributed as follows:

$$\Gamma = 6A_g + 3B_{1g} + 6B_{2g} + 3B_{3g} \qquad (1)$$

Because ZrAs$_2$ has a centrosymmetric structure, only even symmetry modes are allowed in Raman spectra. The Raman tensors for these modes are:

$$A_g = \begin{pmatrix} a & 0 & 0 \\ 0 & b & 0 \\ 0 & 0 & c \end{pmatrix} \quad B_{1g} = \begin{pmatrix} 0 & d & 0 \\ d & 0 & 0 \\ 0 & 0 & 0 \end{pmatrix} \quad B_{2g} = \begin{pmatrix} 0 & 0 & e \\ 0 & 0 & 0 \\ e & 0 & 0 \end{pmatrix} \quad B_{3g} = \begin{pmatrix} 0 & 0 & 0 \\ 0 & 0 & f \\ 0 & f & 0 \end{pmatrix} \qquad (2)$$

Figure 3 presents room temperature Raman spectra recorded at the (YY) geometry for three excitation energies: 2.62 eV (473 nm), 1.96 eV (633 nm), and 1.58 eV (785 nm) with continuous background dependent on excitation energy. The background in spectra will be discussed further below. Phonon modes positions, linewidths and intensities were analyzed after background subtraction. Generally, the Raman spectra measured in this study are consistent with selection rules stemming from the Raman tensor with some "leaking" effect due to resonant conditions. Apparent difference is observed for the lowest frequency $A_g$ mode at 94.5 cm$^{-1}$, which has a significant intensity in the forbidden cross polarizer configuration (Fig. 4).

Table 1 gives the assignment of the observed phonon modes according to their symmetry compared with calculated frequencies (VASP_FP). The polarized angular-resolved Raman spectra confirm this assignment (Fig.S1. in the Supplement presents the polar graphs for selected modes, as a function of the θ angle between the light polarization vector and the b-axis.

Figure 5 shows spectra for (YY) and (X'X') polarization for different excitation wavelengths. Intensities of several Raman peaks have distinctive excitation wavelength dependence. The $A_g$ symmetry modes at 129 cm$^{-1}$ and 171 cm$^{-1}$ have much higher intensity for the 633 and 785 nm excitations than for the 473 nm excitation. The $A_g$ mode at 276 cm$^{-1}$ does not appear in the 473 nm spectrum. Mode at 244 cm$^{-1}$ ($B_{2g}$) is the most intense for the 473 nm excitation in the (X'X') configuration and appears as a small kink in the 785 nm plot (it is not present in the 633 nm plot).

Usually, in resonant Raman spectra, all modes are seen to increase their intensity when the excitation energy is close to a characteristic electronic transition (Van Hove singularity or band nesting). In the Raman spectra of ZrAs$_2$, we observe both resonance and antiresonance effects depending on mode. This means that different electronic intermediate states are involved in the Raman scattering. Since in the band structure of ZrAs$_2$ (Fig. 6a), many bands are available for such transitions, different bands in the whole Brillouin zone take part in the light scattering process. It leads to a quantum interference effect, when electronic transitions in different parts of the Brillouin zone enhance or quench each other [21-24]. Due to the complicated band structure of ZrAs$_2$, we cannot identify the regions that contribute constructively to the Raman amplitude and those that interfere destructively. It is worth noting that differences in electron-phonon matrix elements for phonon modes can also account for observed excitation dependence [24].

As shown in Fig. 3, the continuous background is present in the ZrAs$_2$ Raman scattering spectra. The background shows a quasi-elastic scattering (QES) wing and a flat finite energy continuum extending up to 1500 cm$^{-1}$. It is polarization-dependent and has the highest intensity in the (YY) configuration, i.e. for light polarization along the b-axis, where arsenic atoms form covalently bonded chain. It also depends on excitation energy and has the highest intensity with an intense QES part for the lowest used excitation energy of 1.58 eV (785 nm). For the higher excitation energies, 1.96 eV and 2.62 eV, the intensity of the quasi-elastic component is weaker, and the broad energy range background is observed. The background intensity measured with the 785 nm excitation decreases with increasing temperature (inset in Fig. 3).

A broad band of continuous energy Raman scattering was observed for some topological semimetals: e.g. in Cd$_3$As$_2$ [7,25], WP$_2$ [26], LaAlSi [21]. It is attributed to electronic Raman scattering (ERS) due to electron density fluctuations. Such a background was first observed in heavily doped semiconductors [27, 28]. It was interpreted as scattering by intervalley density fluctuations in the collision-limited regime [29]. Zawadowski and Cardona [30] have shown, that in Raman spectra of metals a background can be due to a scattering between different parts of Fermi surface. Intense QES is also observed in the metallic phase of La$_{0.7}$Sr$_{0.3}$MnO$_3$ [31]. The quasi-elastic part of the continuum background is often described by the Lorentzian profile for zero energy:

$$I_R \propto \left(n(\omega)+1\right)\frac{\omega \Gamma_{el}}{\omega^2 + \Gamma_{el}^2} \qquad (3)$$

where $n(\omega)$ is Bose factor, $\Gamma_{el}$ is the width of the Lorentz profile.

Equation (3) gives a good fit to the QES wing in the 785 nm excitation spectra (Fig. 3). The intensity of this background decreases with temperature in the range of 80 – 443 K (inset in Fig. 3). $\Gamma_{el}$ is almost independent of temperature and has an average value of $(230 \pm 20)$ cm$^{-1}$.

Due to the presence of an electronic background in Raman scattering spectra, one can expect effects of interference between discrete phonon modes and electron continuum states resulting in asymmetric Fano-type profiles of Raman peaks intensity:

$$I(\omega) = I_0 \frac{[1 + (\omega - \omega_0)/q\Gamma]^2}{1 + [(\omega - \omega_0)/\Gamma]^2} \qquad (4)$$

where $\Gamma$ is a spectral width, $\omega_0$ is a peak position. The $1/q$ is the asymmetry factor that accounts for the strength of electron-phonon interaction. Identifying the asymmetric Fano shape for closely spaced, partially overlapped peaks was impossible. For all but one of the separated Raman peaks in ZrAs$_2$, symmetric (Lorentz or Gaussian) line shape provided better fits to experimental data than the Fano profile (value of $1/q$ was practically 0). Only for the 171 cm$^{-1}$ A$_g$ mode in (YY) configuration, the peak shape described by the Fano profile was found in the spectra excited by 633 nm and 785 nm lasers. Figure 7 presents comparison of the (YY) and the (X'X') spectra with the fitted Fano profile. In the (YY) configuration the fitted value of asymmetry factor is $1/q$ = -0.19 ± 0.05, and it is nearly independent of temperature. This indicates no significant change in the electronic structure at the Fermi surface as a function of temperature. The (X'X') plot is almost symmetric with $1/q$ value close to zero. It is in keeping with a small relative value of background intensity for the (X'X') spectra compared to the (YY) spectra. The scarcity of Fano-like peaks in Raman spectra of ZrAs$_2$ resembles the situation of another topological semimetal TaAs, where the electron-phonon coupling was identified in temperature evolution of the Raman modes linewidth, but no asymmetric Fano-like Raman peaks were reported [32]. However, asymmetric Fano profiles have been found in the infrared spectra of TaAs [33].

**Effect of temperature on Raman scattering in ZrAs$_2$**

Figure 8 shows the evolution of the Raman spectra measured in the parallel configuration for $\theta = 45°$ in the temperature range 80 K – 443 K for the 2.62 eV (473 nm) excitation. Spectra for the 1.96 eV (633 nm) excitation are shown in Fig.S3 in the Supplement. Most peaks are visible for the whole temperature range. However, some peaks for the 2.62 eV excitation lose their intensity with increasing temperature, e.g. the A$_g$ mode at 280 cm$^{-1}$ (at 80 K) is hardly visible for temperatures above 300 K.

To find the basic parameters of the phonon lines (position, linewidth, area) the fitting procedure has been performed. The best fit was achieved with the pseudo-Voigt function, which is a weighted sum of the Lorentz and Gaussian profiles. Attempts to fit spectra with the Voigt profile give unreliable results because of too low signal-to-noise ratio [34].

The temperature dependence of the optical phonon frequency and linewidth is usually ascribed to two effects: the anharmonic effect due to the phonon-phonon coupling and quasi-harmonic effect due to thermal expansion of the crystal lattice. Anharmonic interaction is analyzed within the extended Klemens model [35, 36], which assumes that phonon decays into two or three acoustic phonons. The change of phonon frequency is given as:

$$\Delta\omega(T) = A\left(1 + \frac{2}{e^x - 1}\right) + B\left(1 + \frac{3}{e^y - 1} + \frac{3}{(e^y - 1)^2}\right) \quad (5)$$

Where $\hbar\omega$ is phonon energy, $x = \hbar\omega/2k_BT$, $y = \hbar\omega/3k_BT$ and A and B denote anharmonic constants related to three phonon processes (decay of optical phonon into two phonons) and four phonon processes, respectively. Similar expression describes the temperature evolution of the phonon linewidth due to the anharmonic interaction:

$$\Gamma_{anh}(T) = C\left(1 + \frac{2}{e^x - 1}\right) + D\left(1 + \frac{3}{e^y - 1} + \frac{3}{(e^y - 1)^2}\right) \quad (6)$$

The thermal expansion contribution to frequency change in the linear approximation is given as:

$$\omega(T) = \omega_0 + \chi T \quad (7)$$

Figure 9 presents the temperature dependence of the position and the linewidth of several phonon peaks in ZrAs$_2$. Results for the 473 and 633 nm excitation agree well. In the temperature range used in our experiment, the redshift of phonon frequency is for most modes close to the linear formula (7), with slight deviation at temperatures near 80 K. Fitting with the anharmonic expression (5) also produces almost straight line, so these effects are nearly indistinguishable. However, for the modes at 166 cm$^{-1}$ and 245 cm$^{-1}$, the temperature redshift of phonon frequency is distinctly nonlinear and follows the expression (5).

The temperature dependence of the linewidths of Raman peaks is often successfully described by an anharmonic expression (6). It predicts an increase in linewidth with increasing temperature. However, several phonon modes in Raman spectra show a reduction in linewidth for increasing temperature (Fig. 9). It is due to important contribution of the electron-phonon coupling, which is particularly important in semimetals. In this interaction phonons decay into electron-hole pairs via intra- or inter-band transitions close to the Fermi level ($E_F$) [8,37,38]. Temperature dependence of the linewidth is determined by the difference in occupation of electronic states below and above $E_F$. For increasing temperature, the occupation of the electron states below $E_F$ decreases, while the occupation of the states above $E_F$ increases. It leads to decreasing number of available electronic states for the phonon induced transitions and results in a decrease of the linewidth with increasing temperature. This behavior is quantitatively expressed by the formula [37, 38]:

$$\Gamma_{eph} = \gamma_{eph}\left[\left(f(-\hbar\omega/2)\right)-\left(f(\hbar\omega/2)\right)\right] \quad (8)$$

where $f(\hbar\omega) = \left(\exp(\hbar\omega/k_B T)+1\right)^{-1}$ is the Fermi-Dirac distribution function,

It is worth noting that taking into account of a finite chemical potential in (8) can result in a nonmonotonic temperature dependence of $\Gamma_{eph}$ [8, 32].

The decay of optical phonon with zero wavevectors via the creation of the electron-hole pairs depends on the energy and symmetry of phonon mode. Since the highest energy phonons at the center of the Brillouin zone of $ZrAs_2$ have an energy of 34 meV, electron-phonon coupling is possible for pairs of electron bands below and above the Fermi energy, which are closer to each other than 34 meV. It can happen only with an electron k-vector along the Γ - Z line and around the T point in the Brillouin zone (Fig. 6c). However, symmetry-based selection rules (with and without spin-orbit coupling) do not allow Raman active phonons to induce interband transitions at the T point. Due to spin-orbit coupling, such transitions are allowed near the T point and along the Γ - Z line. We cannot exclude that conditions for effective electron-phonon coupling may also exist at general k-points in the Brillouin zone.

In an analysis of the temperature dependence of the linewidth, we used an expression with two contributions: anharmonic term $\Gamma_{anh}$ - equation (6) (we put $D = 0$, since four phonon processes give negligible contribution), and the electron-phonon coupling term $\Gamma_{eph}$ according to the expression (8).

$$\Gamma = \Gamma_0 + \Gamma_{anh} + \Gamma_{eph} \quad (9)$$

We investigated the temperature dependence of the linewidth of phonon peaks represented by the full width at half maximum (FWHM) for two excitation energies, 1.96 eV (633 nm) and 2.61 eV (473 nm), whenever the peak was present and measurable in the spectra for a given excitation energy. For low intensity modes and overlapping peaks it was impossible to get reliable values of the linewidth. Temperature dependence of the linewidth measured for the two excitations exhibits significant differences (Fig. 9). In the spectra recorded with 633 nm laser we observe for most modes almost linear linewidth increase in accord with anharmonic model, apart from the 94.5 cm$^{-1}$ mode, when the 633 nm plot follows the 473 nm line. It is unclear to us what the source of these discrepancies is; one of the possible reasons can be a difference in light penetration depth for the used excitation energies. For the 473 nm excitation, the contribution of the electron-phonon interaction to phonon decay is seen for several modes as a monotonous decrease or a minimum in the temperature dependence of the linewidth. This type of temperature dependence is observed for the modes of A$_g$ symmetry at 94.5 cm$^{-1}$, 129 cm$^{-1}$, and 223 cm$^{-1}$, for two modes of B$_{2g}$ symmetry at 150 and 245 cm$^{-1}$, and the 166 cm$^{-1}$ B$_{3g}$ mode. The dominance of the electron-phonon coupling over anharmonic decay is observed for the modes at 94.5 and 166 cm$^{-1}$, where no contribution from the anharmonic term is needed (C = 0) to fit the experimental data. This observation indicates no clear-cut correlation between the strength of the electron-phonon coupling in phonon decay, and phonon symmetry or frequency. It is an individual property of the phonon mode.

The values of the $\gamma_{eph}$ parameter, characterizing the strength of electron-phonon coupling for phonon modes (Table 2), are smaller than the values reported for some phonon modes in NiTe$_2$ - 5.41 cm$^{-1}$ [39] and PdTe$_2$ - 28.8 cm$^{-1}$ [40] as well as in graphene and graphite - appr. 10 cm$^{-1}$ [38]. Still, the electron-phonon coupling plays a significant role in phonon decay processes in ZrAs$_2$.

CONCLUSIONS

We have investigated Raman scattering in a Dirac nodal line semimetal ZrAs$_2$. Raman spectra have been recorded for several excitation laser energies at different light polarizations. The calculated zero wavevector phonon frequencies and polarization dependence of the Raman peaks enabled symmetry identification of all observed phonon modes. Due to an interference between excitation paths with different intermediate states, significant peak intensity differences exist for different excitation wavelengths. The polarization-dependent electronic background is present in the Raman spectra, with an intense quasi-elastic scattering in the spectra recorded with the 785 nm excitation. The asymmetric Fano peak is observed for the 171 cm$^{-1}$ phonon mode due to interference of a phonon with intense electronic background observed for light polarization along the b axis of the ZrAs$_2$. Effects of electron-phonon interaction manifest themselves in decreasing peak linewidth with increasing temperature for modes of different symmetry, indicating differences in the electron-phonon coupling. We identified points in the Brillouin zone at the Fermi level where optical phonons can decay via e-h pair creation: only for the electronic states on the Γ - Z high symmetry line and in the vicinity of the T point, phonons are allowed to create e-h pairs. Since in ZrAs$_2$ there are no Dirac points close to the Fermi energy, it is hard to expect a significant influence of these points on the electron-phonon coupling.

ACKNOWLEDGEMENTS

This research was partially supported by the Foundation for Polish Science, facilitated by the IRA Programme and co-financed by the European Union within the framework of the Smart Growth Operational Programme (Grant No. MAB/2017/1).

# Tables

**Table 1.** Symmetry assignment of the observed Raman modes in ZrAs$_2$. The mode frequencies were determined from the room temperature spectra. Accuracy of mode frequency values is appr. 1 cm$^{-1}$. Theoretical values were calculated using the DFT method.

|   | Frequency (cm$^{-1}$) | | Symmetry assignment |
|---|---|---|---|
|   | Experiment | Theory (VASP FP) |   |
| 1  | 94.5  | 87.8  | $A_g$ |
| 2  | -     | 89.8  | $B_{1g}$ |
| 3  | 129   | 121.9 | $A_g$ |
| 4  | 140   | 134.5 | $B_{3g}$ |
| 5  | 149   | 137.3 | $B_{2g}$ |
| 6  | 166   | 154.2 | $B_{3g}$ |
| 7  | 171   | 161.2 | $A_g$ |
| 8  | 173.5 | 164.4 | $B_{2g}$ |
| 9  | 178   | 170.1 | $B_{1g}$ |
| 10 | 182.5 | 173.6 | $B_{2g}$ |
| 11 | 197   | 186.5 | $A_g$ |
| 12 | 203   | 193.9 | $B_{3g}$ |
| 13 | 207   | 195.3 | $B_{2g}$ |
| 14 | 217   | 206.4 | $B_{1g}$ |
| 15 | 223   | 210.9 | $A_g$ |
| 16 | -     | 211.7 | $B_{2g}$ |
| 17 | 244   | 232.2 | $B_{2g}$ |
| 18 | 276   | 261.0 | $A_g$ |

**Table 2.** Fitting parameters of the temperature dependence of peak position: $\chi$ coefficient equation (7), and linewidth: anharmonic parameter $C$ (6), electron-phonon coupling parameter $\gamma_{eph}$ (8) for the 473 nm excitation. Additionally, for the 94.5 cm$^{-1}$ mode parameters from the fit for the 633 nm excitation spectra are given.

| Mode frequency (cm$^{-1}$) | Symmetry | $\chi$(cm$^{-1}$/K) | C(cm$^{-1}$) | $\gamma_{eph}$(cm$^{-1}$) |
|---|---|---|---|---|
| 94.5 (473 nm) | A$_g$ | -0.0064 | 0 | 1.86 ± 0.03 |
| 94.5 (633 nm) | A$_g$ | -0.0064 | 0 | 1.68 ± 0.02 |
| 129 | A$_g$ | -0.01 | 0.1 ± 0.03 | 1.7 ± 0.5 |
| 166 | B$_{3g}$ | -0.008 | -0.01 | 0.62 ± 0.5 |
| 223 | A$_g$ | -0.018 | 0.21 ± 0.06 | 0.65 ± 0.05 |
| 245 | B$_{2g}$ | -0.017 | 0.32 ± 0.08 | 0.8 ± 0.5 |
| 276 | A$_g$ | -0.022 | 1.36 ± 0.12 | 0 |

Figures

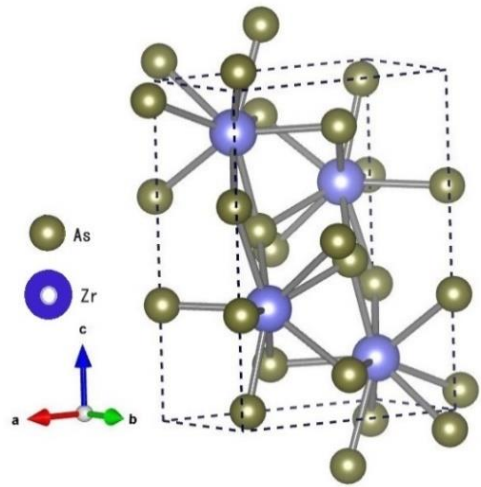

**Fig. 1.** Orthorhombic unit cell of ZrAs$_2$.

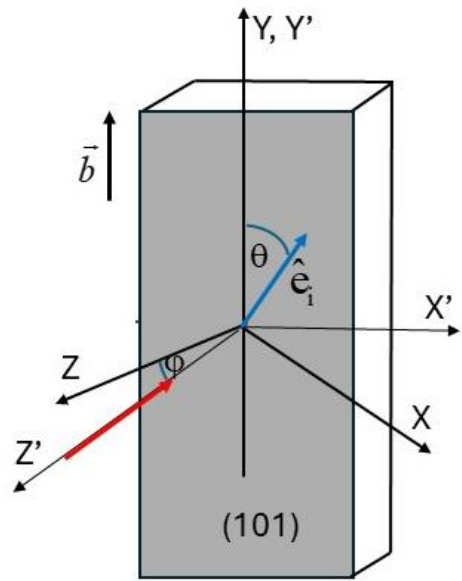

**Fig.2.** Experiment geometry and crystal orientation. The crystal frame of reference (XYZ) and the lab frame of reference (X'Y'Z'). Light polarization (incident and scattered) is in the (X'Y) plane. The red arrow shows the light propagation vector, and the blue denotes the light polarization vector.

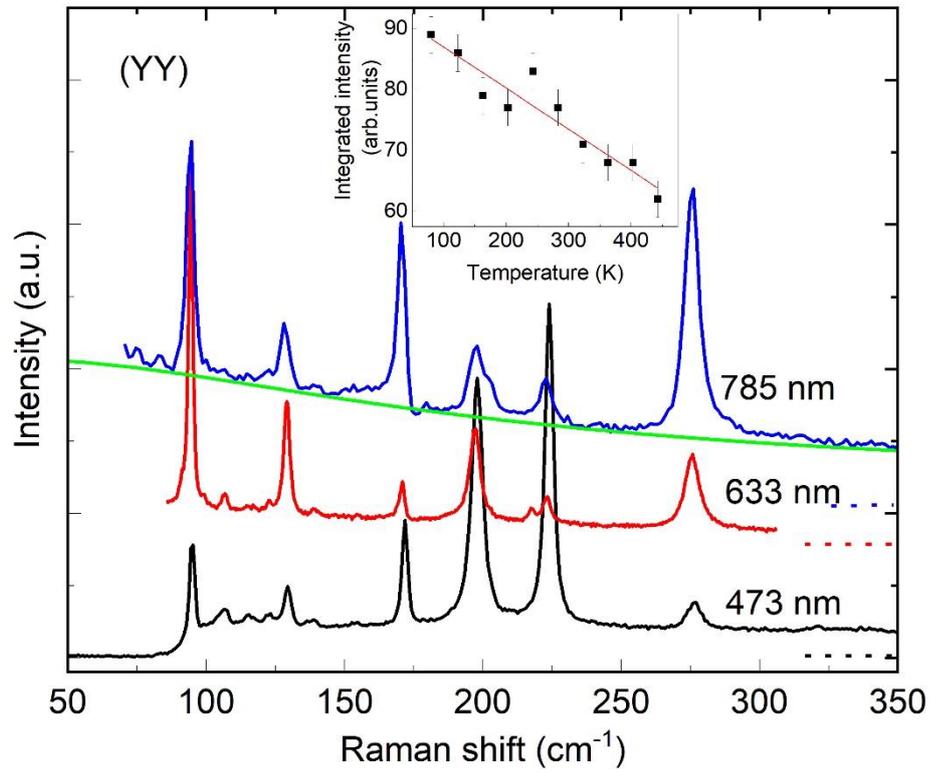

**Fig.3.** Room temperature Raman spectra for three different excitations (as measured). The continuous background is present with the dashed lines as reference levels for every plot. The green line is the fitted model of quasi-elastic scattering background using expression (3) to the 785 nm plot. The inset shows the temperature dependence of the integrated background intensity for the 785 nm excitation.

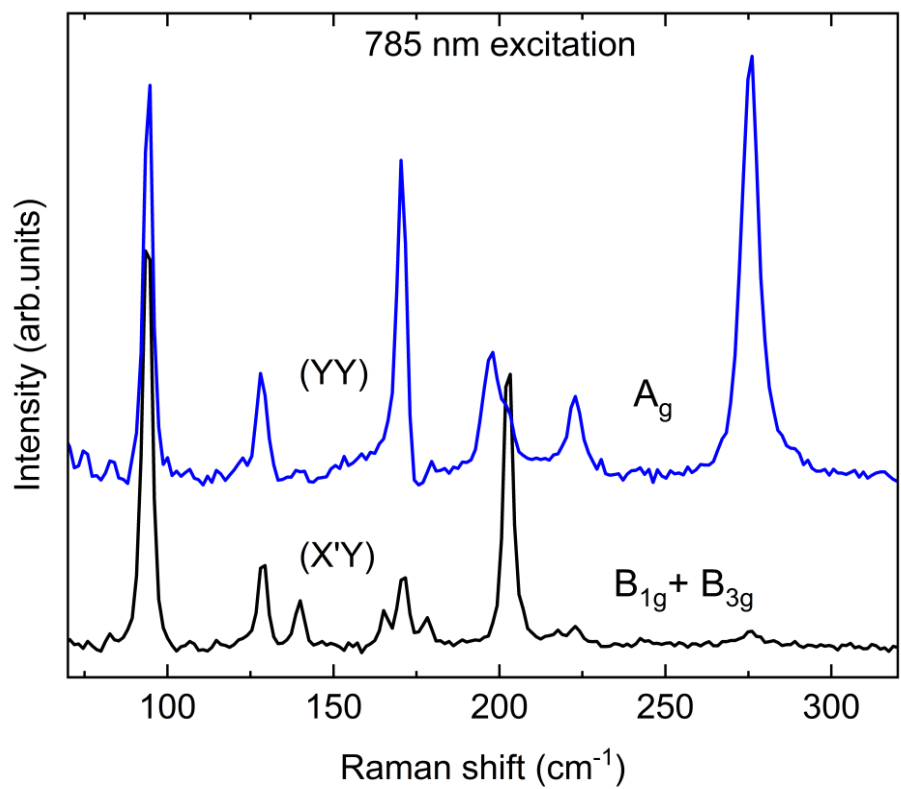

**Fig.4.** Comparison of the parallel (YY) and the (X'Y) crossed polarization spectra recorded with the 785 nm excitation at room temperature.

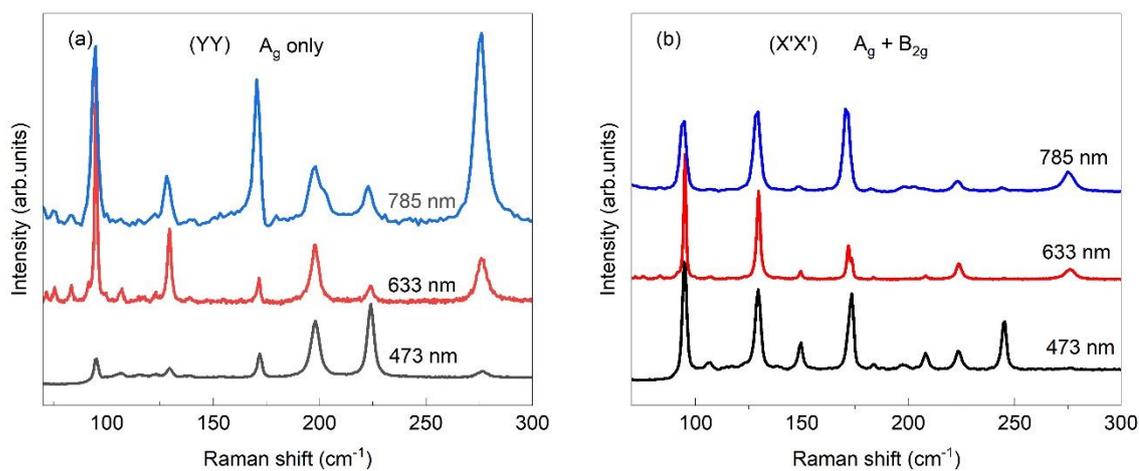

**Fig. 5.** Room temperature Raman spectra in the (YY) (a) and (X'X') (b) configurations for 473 nm, 633 nm and 785 nm excitation wavelengths.

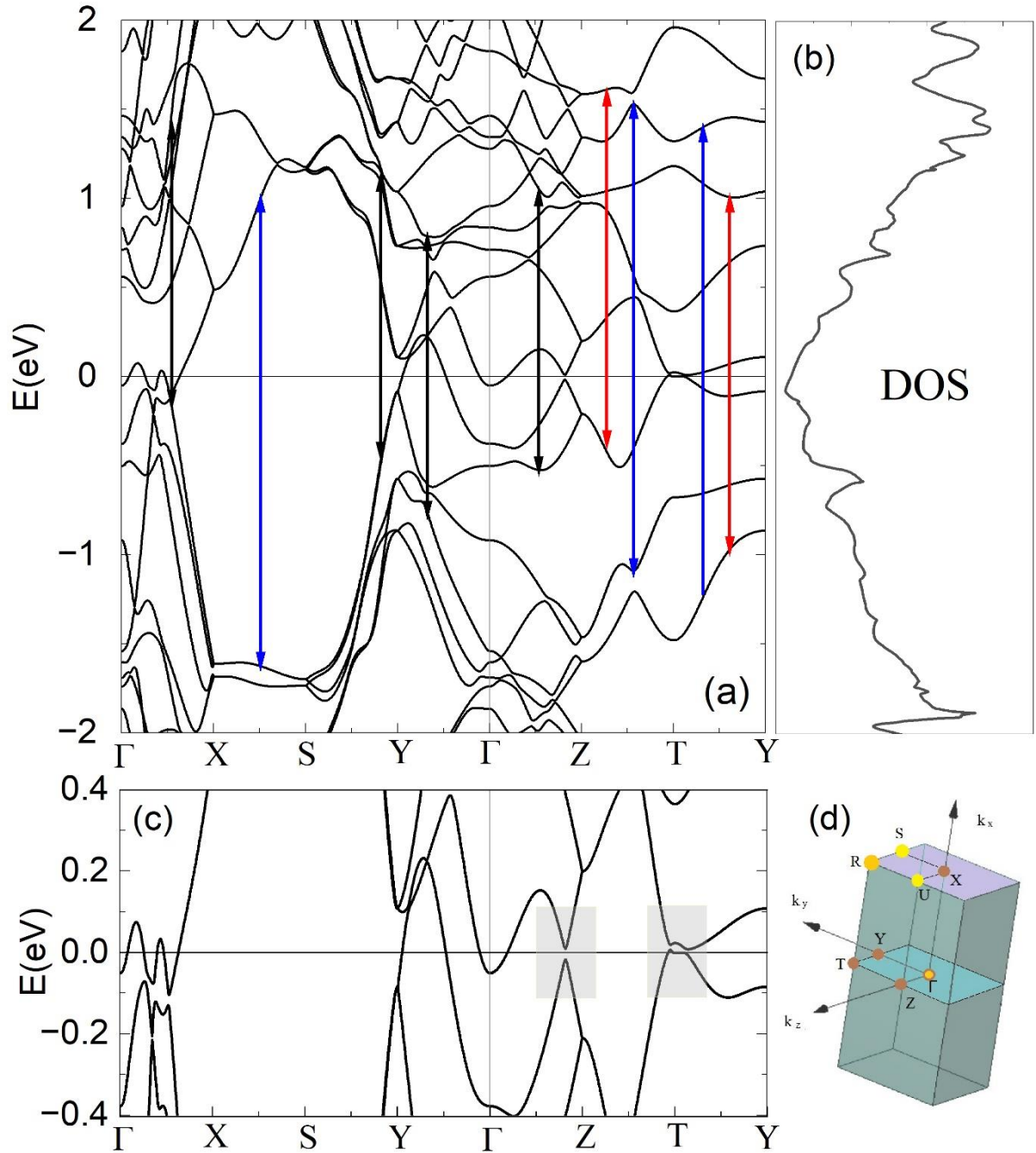

**Fig. 6.** Band structure of ZrAs$_2$ with a few allowed electronic transitions to/from intermediate states in the Raman process, which are shown with the vertical arrows: black - 785 nm excitation, red 633 nm, and blue 473 nm (a); Density of states (b); Details of the electronic bands close to the Fermi level (c); shaded areas indicate regions of the Brillouin zone where phonon decay via e-h pairs excitation is energetically possible; Brillouin zone of ZrAs$_2$ (d).

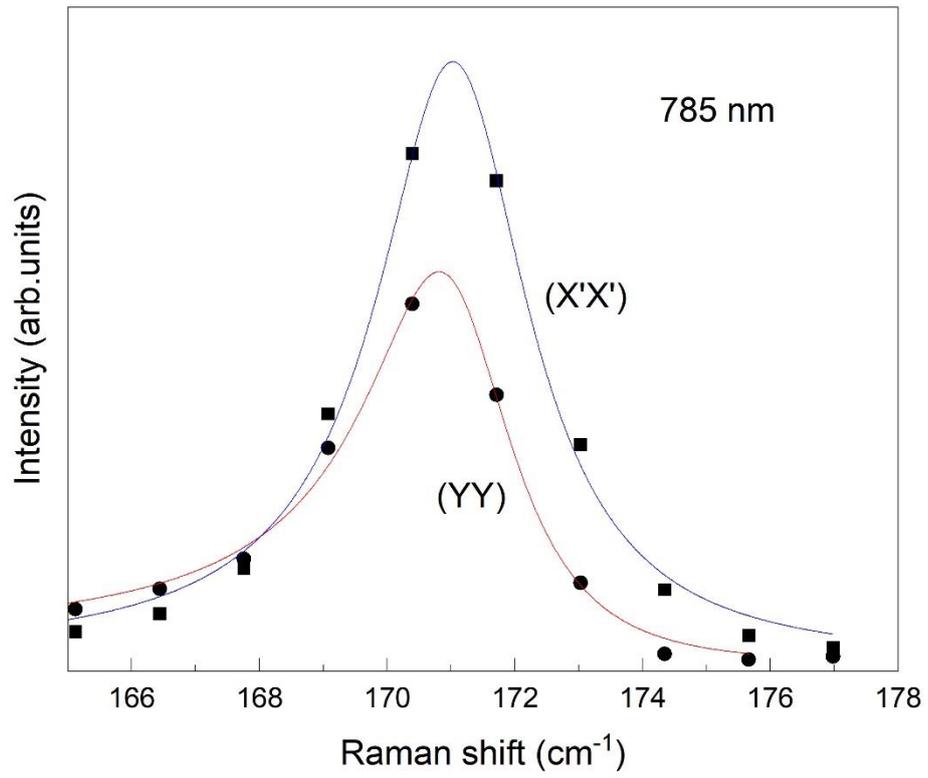

**Fig. 7.** Fano profile fitted to the 171 cm$^{-1}$ A$_g$ phonon peak measured with the 785 nm excitation at room temperature for two parallel light polarizations (YY) and (X'X').

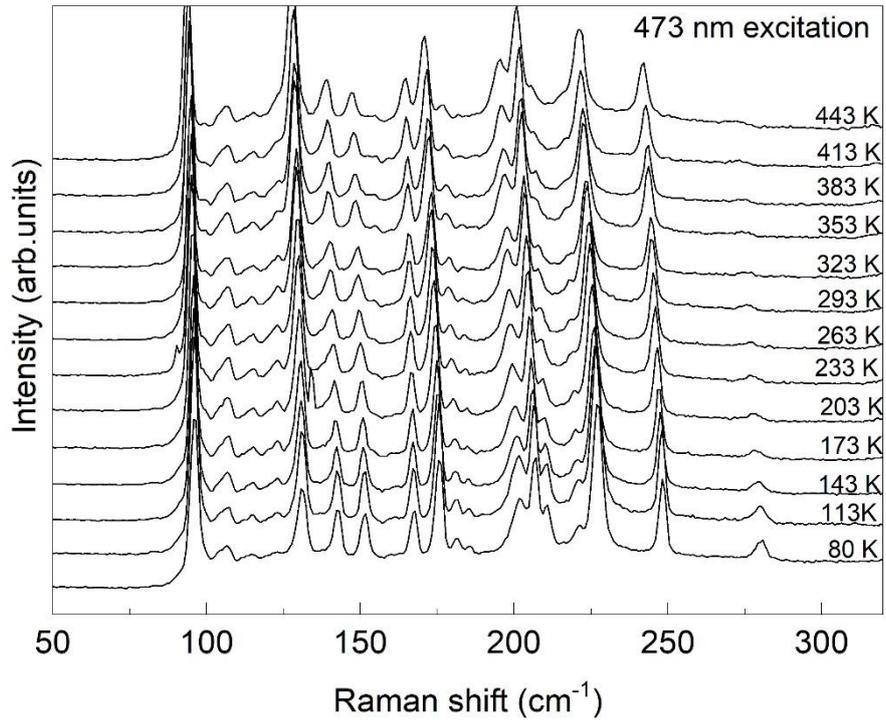

Fig. 8. Temperature dependence of the Raman spectra measured with the 473 nm excitation for parallel configuration and θ = 45° (for the 633 nm excitation see the Supplement).

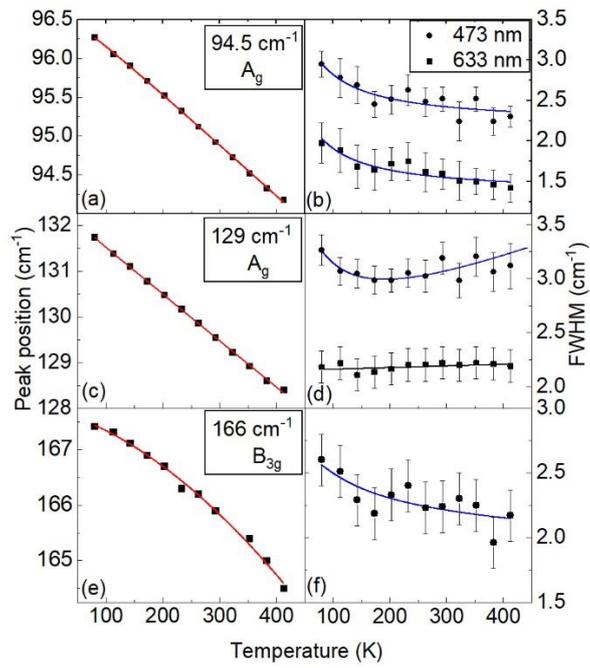
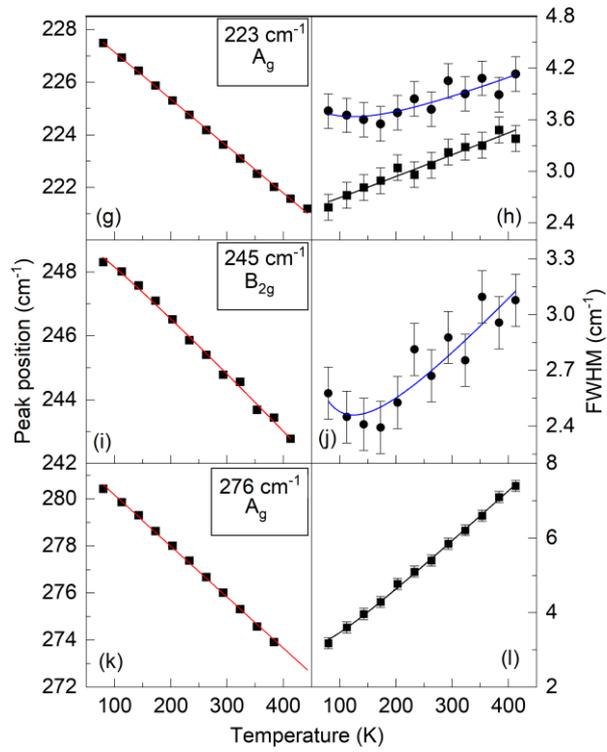

**Fig.9.** Temperature dependence of phonon modes frequency in ZrAs$_2$ (left panels) and the linewidth (right panels) for two excitation wavelengths: 473 nm and 633 nm. The red line is a fit of the expression (5) to the mode frequency change. At the right panels, blue line is a fit of the model where both anharmonic effects and electron-phonon interaction are included. The black line corresponds to the linewidth model of anharmonic interaction only (6). The linewidth results were not corrected for the instrumental broadening.

# SUPPORTING INFORMATION

## Temperature and excitation energy dependence of Raman scattering in nodal line Dirac semimetal ZrAs$_2$

*R. Bacewicz[1], C. Jastrzębski[1], K. Zberecki[1], A. S. Wadge[2], D. Jastrzębski[2] and A. Wiśniewski[2,3]*

[1]*Faculty of Physics, Warsaw University of Technology, Koszykowa 75, PL- 00-662, Warsaw, Poland*
[2]*International Research Centre MagTop, Institute of Physics, Polish Academy of Sciences, Aleja Lotnikow 32/46, PL-02-668 Warsaw, Poland*
[3]*Institute of Physics, Polish Academy of Sciences, Aleja Lotnikow 32/46, PL-02-668 Warsaw, Poland*

**S1. Angular dependence of Raman scattering intensity**

According to the factor group analysis, 18 Raman active modes in ZrAs$_2$ are distributed as follows: $6A_g + 3B_{1g} + 6B_{2g} + 3B_{3g}$. In the identification process of modes symmetry, we measured angular dependence of the modes intensity as a function of the $\theta$ angle (Fig.2 in the main text) in parallel ($\hat{e}_i \parallel \hat{e}_s$) and crossed ($\hat{e}_i \perp \hat{e}_s$) polarization configuration. We expect the following expressions for the intensity of the modes:

A$_g$ modes:

$$I_\parallel \propto \left| b\cos^2\theta + \sin^2\theta\left(a\cos^2\varphi + b\sin^2\varphi\right)\right|^2 \quad \text{(S.1)}$$

$$I_\perp \propto \frac{1}{2}\left|\sin 2\theta(b - a\cos^2\varphi - c\sin^2\varphi\right|^2 \quad \text{(S.2)}$$

B$_{1g}$ modes

$$I_\parallel \propto \left|d\sin 2\theta \cos\varphi\right|^2 \quad \text{(S.3)}$$

$$I_\perp \propto \left|d\cos 2\theta \cos\varphi\right|^2 \quad \text{(S.4)}$$

B$_{2g}$ modes

$$I_\parallel \propto \left|e\sin^2\theta \sin 2\varphi\right|^2 \quad \text{(S.5)}$$

$$I_\perp \propto \left|e\sin\theta \sin 2\varphi\right|^2 \quad \text{(S.6)}$$

B$_{3g}$ modes

$$I_\parallel \propto \left|f\sin 2\theta \sin\varphi\right|^2 \quad \text{(S.7)}$$

$$I_\perp \propto \left|f\cos 2\theta \sin\varphi\right|^2 \quad \text{(S.8)}$$

Where *a, b, c, d, e, f* are the elements of the Raman tensor (see (2) in the main text), $\varphi$ is the angle between Z and Z' axes (Figure 2 in the main text); for ZrAs$_2$ $\varphi$ = 53.0247°.

Fitting the angular dependence of intensity with real elements of Raman tensor fails for $A_g$ modes. To properly describe this dependence, we have to assume complex Raman elements [S1,S2]: $a = |a|e^{i\phi_a}$, $b = |b|e^{i\phi_b}$ $c = |c|e^{i\phi_c}$. It leads to a complicated expression that cannot be deconvoluted to find modules or phase factors. Operationally, it comes down to expressions:

$$I_\parallel = A_1 \cos^4\theta + A_2 \sin^2 2\theta + A_3 \sin^4\theta \quad (S.9)$$

$$I_\perp = A_4 \sin^2 2\theta \quad (S.10)$$

where $A_1$, $A_2$, $A_3$ and $A_4$ are the fitting constants.

For B modes phase factors do not play any role and the real values of Raman tensors are good enough to fit the angular dependence for parallel and perpendicular configurations. Figure S1 shows the angular dependence for six representative modes of different symmetry. The fitted value of the angle offset was 4°.

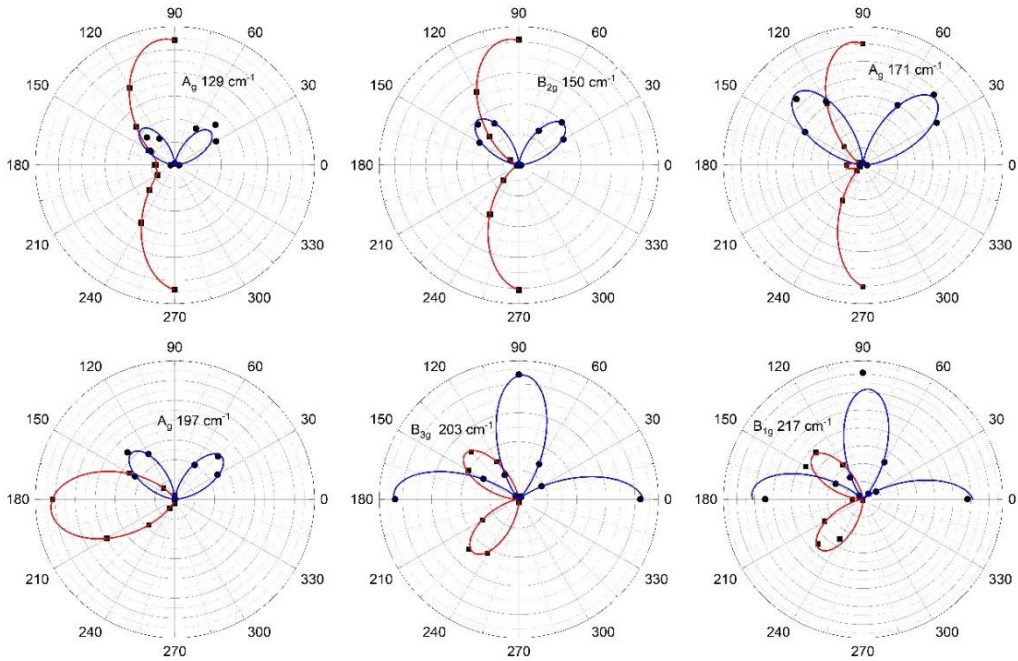

Fig.S1. Polar plots of Raman intensity as a function of the $\theta$ angle between the b-axis of ZrAs$_2$ crystal (it is also growth direction of needle-like crystals) and the incident light polarization vector $\hat{e}_i$ for two configurations parallel $\hat{e}_i \parallel \hat{e}_s$ and perpendicular $\hat{e}_i \perp \hat{e}_s$. Square symbol and red line correspond to parallel configuration, circle symbol and blue line denote the perpendicular configuration.

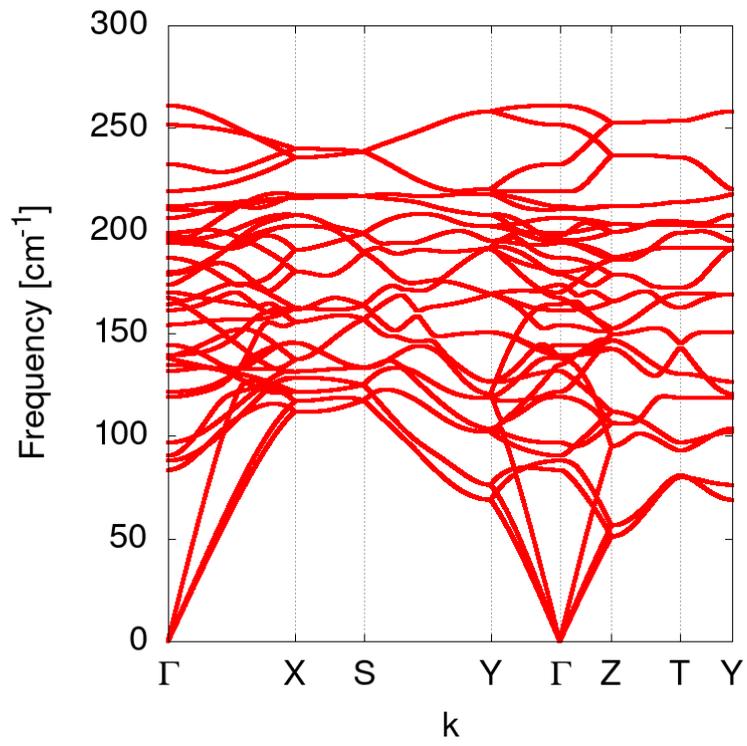

Fig.S2. Phonon dispersion calculated using the frozen phonon method, as implemented in the phonopy code [19].

**S2. Temperature dependence of the Raman spectra for the 633 nm excitation**

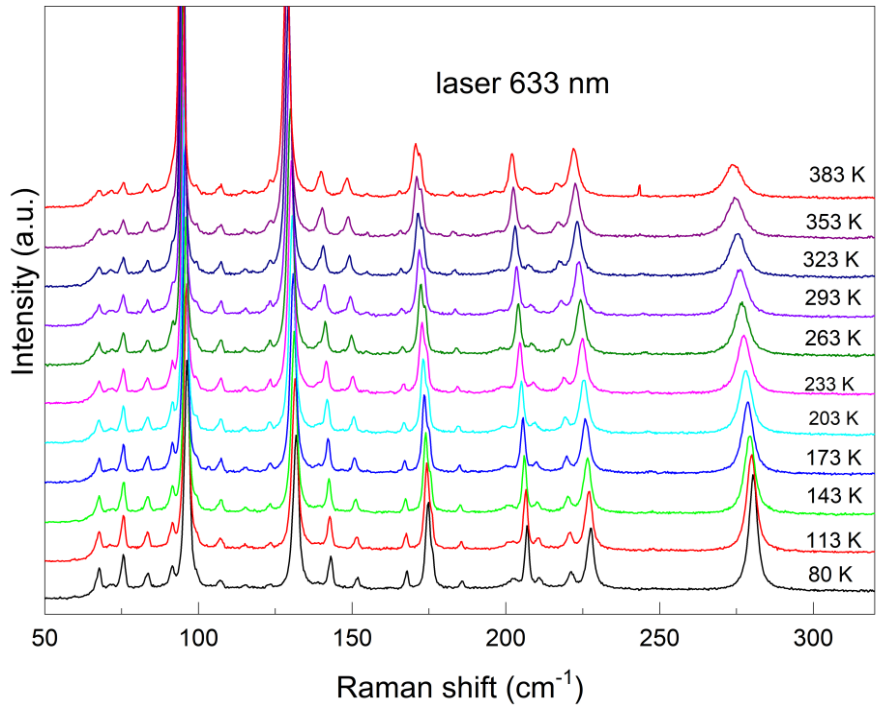

Fig.S3. Temperature dependence of the Raman spectra measured with the 633 nm excitation for parallel polarization configuration with θ = 45°. Periodically distributed peaks below 100 cm$^{-1}$ are due to the air scattering.

REFERENCES

S1. I. Abbasian Shojaei, S. Pournia, C. Le, B. R. Ortiz, G. Jnawali, Fu-Ch. Zhang, S. D. Wilson, H. E. Jackson, and L. M. Smith, *A Raman probe of phonons and electron–phonon interactions in the Weyl semimetal NbIrTe$_4$* Sci. Rep. **11**, 8155 (2021)

S2 H.B. Ribeiro, M.A. Pimento, C.J.S. de Matos, R.L. Moreira, A.S. Rodin, J.D. Zapata, E.A.T. de Souza, and A.H.Castro Neto, *Unusual Angular Dependence of the Raman Response in Black Phosphorus* ACS Nano **9**, 4276 (2015)